# GLUON VERSUS MESON EXCHANGE IN HADRON-HADRON SYSTEMS ON THE LATTICE [1]

H. MARKUM, K. RABITSCH, W. SAKULER

*Institut für Kernphysik, Technische Universität Wien*
*A-1040 Vienna, Austria*

The interaction of spatially extended heavy hadrons is investigated in the framework of lattice QCD with dynamical quarks. In addition to the baryon-baryon potential results for the baryon-antibaryon and for the meson-meson system are presented. It is shown that the expected dipole forces have a very short range and that sea quarks play a minor important role.

[1] Supported in part by "Fonds zur Förderung der wissenschaftlichen Forschung" under Contract No. P7510-TEC.

QCD as a quantum field theory has due to the uncertainty principle a creation probability of virtual gluons and quarks. As a consequence the nucleon-nucleon forces are mediated by gluon exchange between the constituent quarks for short distances whereas for longer distances the production of quark-antiquark pairs is the dominating mechanism [1]. The continuous advance of the techniques of lattice QCD makes the study of systems consisting of a few valence quarks in the presence of sea quarks feasible [2]. This paper investigates the influence of both gluons and dynamical quarks on static hadron-hadron systems. The sea quarks constitute the meson exchange as expected by the Yukawa theory.

In our approximation of QCD the valence quarks of a hadron are restricted to fulfill the static Dirac equation. The gluons are treated as SU(3)-Maxwell fields $U_{x\mu}$ and the sea quarks as Dirac fields $\bar{\psi}_x, \psi_x$ in Kogut-Susskind discretization. The free energy $F(\vec{r}_1, \ldots, \vec{r}_N)$ of a system of $N$ quarks and antiquarks in a gluonic and fermionic field is defined as a thermodynamical expectation value and expressed by the Feynman path-integral for the product of $N$ Polyakov loops $L(\vec{r}_i)$ [3]

$$\begin{aligned}\exp\left(-\frac{1}{T}F(\vec{r}_1, \ldots, \vec{r}_N)\right) &= \frac{\int \mathcal{D}[U]\mathcal{D}[\bar{\psi},\psi]\; L(\vec{r}_1)\ldots L^\dagger(\vec{r}_N)\; e^{-S[U,\bar{\psi},\psi]}}{\int \mathcal{D}[U]\mathcal{D}[\bar{\psi},\psi] e^{-S[U,\bar{\psi},\psi]}} \\ &= \langle L(\vec{r}_1)\ldots L^\dagger(\vec{r}_N)\rangle\;. \end{aligned} \quad (1)$$

$S(U, \bar{\psi}, \psi) = S_G(U) + S_F(U, \bar{\psi}, \psi)$ is the total action of the system with temperature $T$. For the gluonic action $S_G$ we use Wilson's plaquette action with the inverse gluon coupling $\beta = 6/g^2$ and for the fermionic action $S_F$ we employ the Kogut-Susskind formulation with $n_f$ flavors. The time evolution of the valence quarks is taken as the static quark propagator, the so-called Polyakov loop $L(\vec{r}) = 1/3\; \text{Tr} \prod_{i=1}^{N_t} U_{\mu=4}(\vec{r}, t_i)$. We construct a quark wave function $\Psi(\vec{r})$ by distributing the static quark charges in Gaussians

$$\Psi(\vec{r}) = (\sqrt{\pi}\sigma)^{-3/2}\; e^{-\frac{r^2}{2\sigma^2}}\; L(\vec{r}) \quad (2)$$

over a sphere with radius $R$ on the lattice. Since the results turn out to depend only weakly on the width $\sigma$ we set (2) to a uniform distribution. For the description of baryons we take the product of three quark wave functions and for mesons correlated wave functions of quark and antiquark. Because the static quarks carry no spin and flavor all baryons and mesons are degenerate. The total wave function of the hadron-hadron system is constructed to be a product of Gauss functions with two centers separated by some distance $d$.

To compute the path integral (1) we perform 10000 Monte Carlo iterations for QCD without dynamical fermions and 1000 iterations in the presence of fermions. We take a space-time lattice of size $8 \times 8 \times 16 \times 4$ with periodic boundary conditions. In the pure gluonic case we choose $\beta = 5.6$ which corresponds via the renormalization group equation to a lattice constant of $a \approx 0.25$ fm and to a spatial extension in $z$ direction of about 2 fm. For full QCD at $\beta = 5.2$ ($\beta = 4.9$) we set the number of flavors $n_f = 3$ ($n_f = 4$) and the dynamical quark mass $m = 0.1$. This choice leads to comparable lattice constants all lying in the confinement regime.

We calculate the total energy of the hadron-hadron system according to (1) and subtract twice the energy of a single hadron. In Fig. 1a we show the relative potential energy of the two-baryon system with baryons of radius $R = 1$ for varying distances $d$ between the two clusters. We find an attractive potential with an effective range of about $2R$. The reason for the attractive forces between the two baryons is the Coulomb plus linear type of potential between quarks which tries to attract the two three-quark clusters. When the two baryons become separated their colors are saturated and they evolve to isolated color singlets which do not interact. At least on the lattice no long range forces can be resolved. Taking dynamical fermions into account we obtain no drastic effect. This is also observed in mass spectrum calculations where sea quarks give rise to only a 10 per cent effect. In Fig. 1b we display the two-baryon system with baryon radius $R = 2$. We find attractive interactions only in the overlap region and practically no flavor dependence.

We now turn to the baryon-antibaryon system in Fig. 1c and again see an attractive interaction in the overlap regime. The difference between the baryon-baryon and baryon-antibaryon potentials might be interpreted qualitatively by the fact that the baryon-antibaryon system can build three mesons which is energetically favored. Switching on dynamical fermions we observe practically no effect. Increasing the radius of the hadron sphere to $R = 2$ we recognize in Fig. 1d more or less the same behavior as for $R = 1$. The static meson-meson system is depicted in Figs. 1 e–f for radii $R = 1$ and $R = 2$, respectively, and behaves similar to the previous systems.

To conclude, this study deals with a truncated Dirac equation restricting the valence quarks to static color charges. The construction of the quark orbitals leads to baryons of reasonable size but takes no kinetic energy into account. It turns out that the gluon exchange dominates and sea quark effects are small. The aim would be to evaluate the $S$ matrix for scattering of two nucleons consisting of three quarks each. This requires the calculation of a four-point hadron-Green-function being equivalent to a correlation function of six quark propagators in the QCD path-integral. There has been some progress in computing meson scattering-amplitudes from the finite volume dependence of the bound-state energies [4]. A method to extract effective potentials from the two-hadron spectrum was proposed recently [5].

# References

[1] K. Holinde, *Phys. Rep.* **68** (1981) 121.

[2] H. Markum, M. Meinhart, G. Eder, M. Faber and H. Leeb, *Phys. Rev.* **D31** (1985) 2029; K. Rabitsch, H. Markum and W. Sakuler, *Phys. Lett.* **B**, in print.

[3] L. D. McLerran and B. Svetitsky, *Phys. Rev.* **D24** (1981) 450.

[4] S. R. Sharpe, R. Gupta and G. W. Kilcup, *Nucl. Phys.* **B383** (1992) 309.

[5] H. R. Fiebig, Talk at the 3$^{\text{rd}}$ Workshop on Lattice Field Theory (Vienna, 1993).

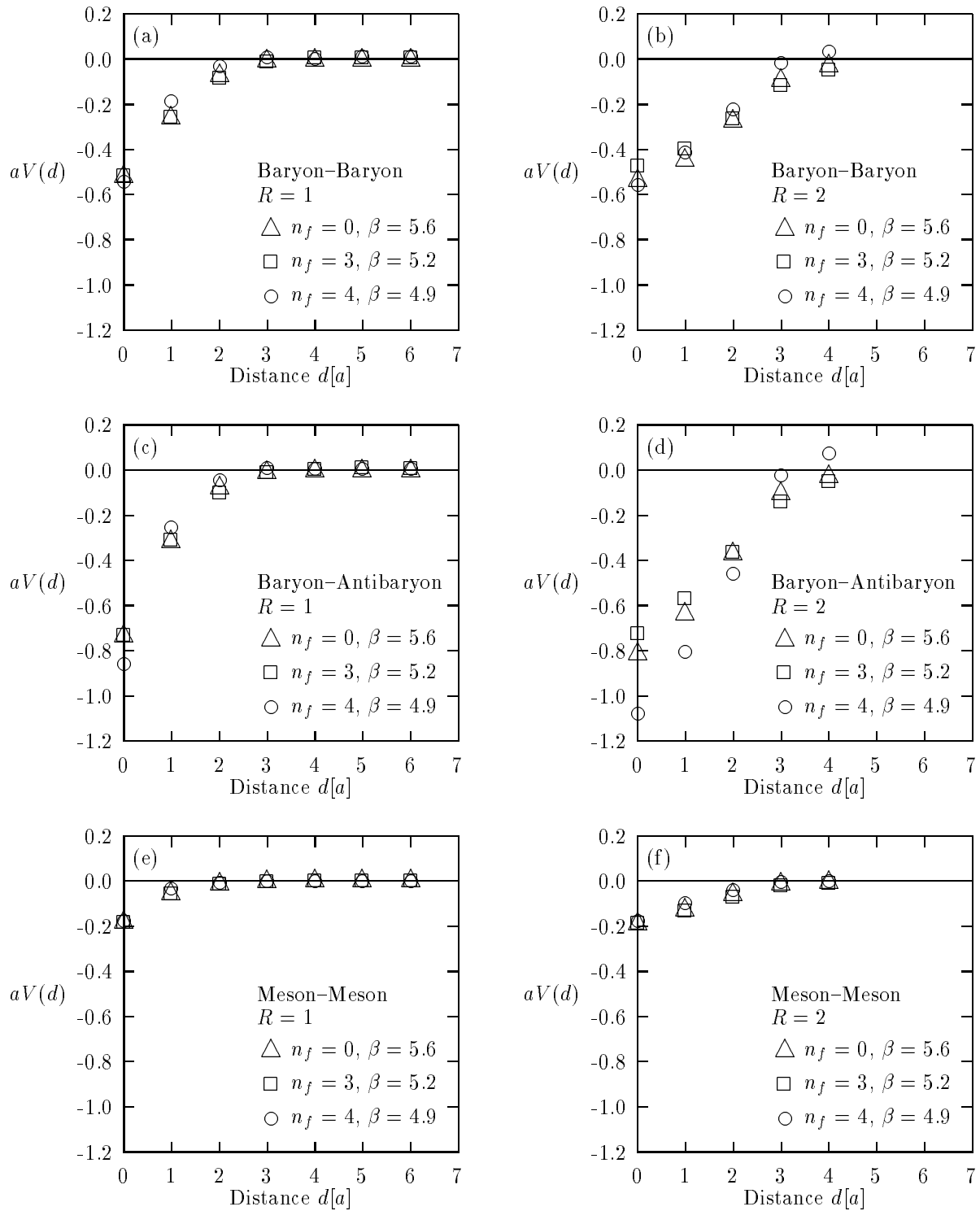

Fig. 1: Baryon-baryon potentials (a,b) as a function of center of mass distance $d$ for pure gluon exchange and with sea quarks for baryons with radii $R = 1$ and $R = 2$ in lattice units. The interaction is attractive and takes place mainly in the overlap region. The baryon-antibaryon potentials (c,d) turn out to be deeper and the meson-meson potentials (e,f) are shallower. Dynamical quarks have no pronounced effect. Error bars are in the size of the symbols.

5